# Increasing the failure recovery probability of atomic replacement approaches

H. Saboohi[a*], S. Abdul Kareem[a]

[a]Department of Artificial Intelligence, Faculty of Computer Science and Information Technology,
University of Malaya, 50603 Kuala Lumpur, Malaysia
*<saboohi@siswa.um.edu.my>

**Abstract:** Web processes are made up of services as their units of functionality. The services are represented as a graph and compose a synergy of service. The composite service is prone to failure due to various causes. However, the end-user should receive a smooth and non-interrupted execution. Atomic replacement of a failed Web service to recover the system is a straightforward approach. Nevertheless, finding a similar service is not reliable. In order to increase the probability of the recovery of a failed composite service, a set of services is replaced with another similar set.

## 1. Introduction

Composite Web services enable Web processes to fulfill the users' complex requests. The users expect a smooth execution of the service they request. However, it is an inescapable truth that some Web services may fail during execution. The failures are caused by the provider of the service that wants to alter the business logics or the infrastructure that supplies the access.

A straightforward approach to recover the system is to replace the failed service with a similar service [1]. It has been proposed to replicate each Web service so that, if a network problem occurs, for example, the replicated service is executed in lieu of the failed service. However, the failures caused by business changes are not recoverable even with a replicated service. Moreover, there are some so-called *critical services* that it is not possible to replicate. If failover services do not respond there should be other ways to hinder the system from a disruption.

## 2. Subdigraph Renovation

Many existing efforts substitute a failed service with another composite Web service. In accordance to the atomic replacement, we call these approaches "atomic-to-composite replacement".

A movement forward is to apply a "composite replacement" in which a set of services, including the failed service, is replaced with another similar set of services. By "similar," we mean that their functionalities are equal and non-functional properties such as execution time and cost are close enough.

We proposed a graph-based approach in [2] to recover service failures by eliminating a sub-graph of a composite service and inserting a new sub-graph. The proposed approach rests on a design-time phase and a run-time phase. The basic idea is that the design phase anticipates the possible fragmentations of a graph representing the behavior of a Web service and their replacements. The run-time phase, based on the information gathered, does the actual replacement following a failure.

The separation to offline and online phases makes the recovery faster. The time consuming calculations are done before commencing the execution. Therefore, at failure time the replacement is done with no delay.

## 3. Evaluation

An obstacle to evaluating composite Web services is a lack of a standard test collection. Current test collections such as OWLS-TC only contain atomic services.

In order to show the unreliability of the atomic replacement approach we synthetically created a test collection of composite services based on OWLS-TC. We created 1000 digraphs of composite services with a combination of orders between 2 and 6.

We simulated the execution of the composite services and declared the failure of each constituent service. Finally, we investigated whether an atomic service was available to replace the failed service. Moreover, the sub-graph replacement was applied. We repeated the experiment 100 times and reported the mean values. The results are depicted in Figure 1.

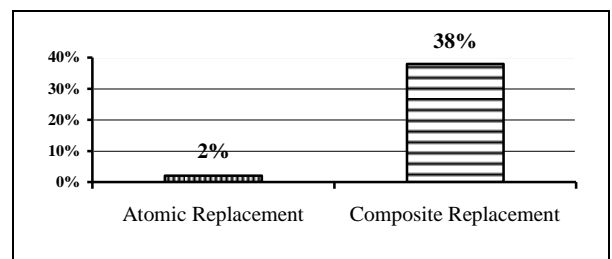

**Figure 1:** Recovery Probability of Atomic and Composite Replacement Approaches

## 4. Conclusion

The results of this research reveal that broadening the replacement fragment from an atomic service to a set of services significantly increases the probability of failure recovery of composite services.